\documentstyle[11pt]{article}
\newcommand{\be}{\begin{equation}}
\newcommand{\ee}{\end{equation}}
\newcommand{\bea}{\begin{eqnarray}}
\newcommand{\eea}{\end{eqnarray}}
\newcommand{\f}{\frac}
\newcommand{\n}{\nabla}

\newcommand{\al}{\alpha}
\newcommand{\bt}{\beta}

\newcommand{\G}{\Gamma}
\newcommand{\vs}[1]{\vspace{#1 mm}}
\newcommand{\hs}[1]{\hspace{#1 mm}}
\newcommand{\lf}{\left(}
\newcommand{\rg}{\right)}

\begin{document}
\hsize=5.6truein
\vsize=9truein
\hoffset=-.3in
\textheight=8.5truein
\voffset=-.8truein
\baselineskip=.60cm
\thispagestyle{empty}
\rightline{hep-th 9905231}

\vs{25}
\centerline{\large\bf On the Geometric Properties of AdS Instantons}
\vs{10}
\centerline{Ali Kaya\footnote{e-mail: ali@rainbow.tamu.edu}}
\vs{5}
\centerline{Center for Theoretical Physics, Texas A\& M University,}
\centerline{College Station, Texas 77843, USA.}
\vs{20}

\begin{abstract}
According to the positive energy conjecture of Horowitz and Myers, there
is a specific supergravity solution, AdS soliton, which has minimum energy
among all asymptotically locally AdS solutions with the same boundary
conditions. Related to the issue of semiclassical stability of AdS
soliton  in the context of pure gravity with a negative cosmological
constant, physical boundary conditions are determined
for an instanton solution which would be responsible for
vacuum decay by barrier penetration. Certain geometric properties of
instantons are studied, using Hermitian differential operators.
On a $d$-dimensional instanton, it is shown that there are $d-2$
harmonic functions. A class of instanton solutions, obeying more
restrictive boundary conditions,  is proved to have $d-1$ Killing
vectors which also commute. All but one of the Killing vectors
are duals of harmonic one-forms, which are gradients
of harmonic functions, and do not have any fixed points.

\end{abstract} 
\vs{50}
\pagebreak


\section{Introduction}


Like for any other field theory, positivity of the energy is a necessary
condition for ensuring the stability of general relativity. In the
presence of gravity, although the notion of local energy density is not
well defined, one can still talk about the total energy of a gravitating
system defined in terms of the asymptotic behavior of the
metric with respect to a background geometry \cite{adm},\cite{ad}. For
asymptotically flat spaces, under general assumptions, the complete proof
of positive energy theorem was first given in \cite{sy}. Later, a simple
and elegant proof was presented in \cite{w1}, using spinors. On the other
hand, it is known that there are non-trivial  zero \cite{w2} and  negative
energy \cite{hd} asymptotically flat spaces on which these proofs do
not  apply.  Generically, such a solution has an asymptotic circle $S^{1}$
which is contractible in the interior, thus the asymptotic geometry  is
only locally Minkowskian. The proof of \cite{sy} does not
apply due to this topological difference, and spinors used in the proof
given in \cite{w1}  are not well defined on such spaces \cite{w2}. This
sector of the theory is completely unstable.\\

Following the remarkable conjecture of 
\cite{m},\cite{gkp},\cite{w}, there is now considerable 
evidence for a correspondence between  certain string theories on
Anti-de-Sitter (AdS) spaces and conformal field theories (CFT)
living on the boundaries of these spaces. For some
applications, the correspondence allows us to work with supergravity
approximations to string theories. Although supersymmetry plays a crucial
role for the conjecture, in \cite{w3}, it has been suggested that 
non-supersymmetric Yang-Mills gauge theory can also be described by
AdS/CFT correspondence by compactifying one
direction on  $S^{1}$ and imposing supersymmetry breaking boundary
conditions to fermions. Recently, Horowitz and Myers have studied the
consequences of this proposal from the point of view of supergravity
theory on AdS \cite{hm}. By AdS/CFT correspondence, one should consider
geometries which have one spatial direction compactified on
$S^{1}$ asymptotically. 
Thus, the asymptotic geometry is not globally but only
locally the AdS space.  The standard techniques can be applied to show the
positivity of energy for spaces which approach globally to AdS at
infinity \cite{gb}. However, the stability of gravity is now questionable
with this  new  asymptotic form, like in the
asymptotically locally Minkowskian sector. Indeed, Horowitz and Myers
presented a solution which is completely regular and has negative total
energy. Following Horowitz and Myers, we will also refer to this solution
as AdS $soliton$. Not surprisingly, 
the topology of AdS soliton is such that
the asymptotic circle at infinity is contractible in the interior.
Horowitz and Myers have beautifully identified the negative energy of this
solution with the negative Casimir energy of the field theory which arises
due to the breaking of supersymmetry. Then, the expected stability of
non-supersymmetric Yang-Mills gauge theory led them to conjecture a new
positive energy theorem for asymptotically locally AdS spaces.\\ 

Perturbative stability of AdS soliton (up to quadratic metric
fluctuations) has been shown in \cite{hm}.  A related and important
problem is the issue of semiclassical stability.
In field theory, a false vacuum decay by barrier penetration 
to a stable ground state via instanton like solutions of Euclidean field
equations, in which at large distances the fields take their values in the
false vacuum \cite{c}. Therefore, one should study the possible
Euclidean metric solutions of Einstein equations which approach the
Euclidean  AdS soliton (which will also be referred as AdS soliton) 
asymptotically. The existence of such a
solution (which will be referred simply as an instanton)
may then imply the instability of the AdS soliton as a result
of a semiclassical analysis (i.e. studying the small fluctuations around
the instanton). With the above motivation,
in this paper,  we will try to determine some geometric properties of
instanton solutions by studying certain differential operators.
Similar arguments were  first used to show
the semiclassical stability of Minkowski space  in \cite{w1}.\\

The organization of the paper is as follows. In section 2, 
we discuss the definition of the Euclidean action, which can be  
expressed as a surface integral when being evaluated on a background, 
and fix the asymptotic behavior of the
metric. In section 3, we introduce two Hermitian 
differential operators defined on divergence free vector fields, and
determine their zero modes. One of the operators has no  zero
modes and turns out to be invertible. We also study the properties of the
Laplacian acting on square integrable functions. 
In section 4, by solving  Dirichlet problems  related to the invertible
differential operators introduced in section 3, we try to prove that  
an instanton has certain number of harmonic functions  and
commuting Killing vectors. 
We solve the Dirichlet problem related to the Laplacian and
show  that on a $d$-dimensional instanton there are $d-2$ harmonic
functions. The Dirichlet problem,  which is intimately
related to the existence of Killing vectors, can be solved in six or
higher dimensional instantons. However, that solution does not have   
appropriate asymptotics to give the desired Killing vectors.
Then, we focus on a family of instantons
obeying more restrictive boundary conditions. On a $d$-dimensional  
instanton of this family, we show that there are $d-1$
commuting Killing vectors. It turns out that all but one of the Killing
vectors are duals of harmonic one-forms, which are gradients 
of harmonic functions, and do not have any fixed points.
We conclude with some brief remarks in section 5.


\section{Euclidean action, surface terms and boundary conditions}


The Lorentzian action of the pure gravity on a $d$-dimensional
spacetime $M_{L}$ with a
negative cosmological constant, $\Lambda <0$, may be
written as:
\be
I_{L}= \f{1}{16\pi G} \int_{M_{L}} (R-2\Lambda ) + 
\f{1}{8\pi G} \int_{\partial M_{L}} K ,
\ee
where $G$ is the gravitational constant, R is the scalar curvature of the
metric $g_{AB}$, $K$ is the  trace of the 
extrinsic curvature of the boundary $\partial
M_{L}$. The volume elements on $M_{L}$ and $\partial M_{L}$ are determined
by $g_{AB}$ and the induced metric on $\partial M_{L}$, respectively,
and will not be written explicitly. The
surface term is added to obtain the correct equations subject only to the
condition that the induced metric on $\partial M_{L}$ is held fixed.\\

Inspired by the Lorentzian form of the action, one may define the
Euclidean action on a manifold $M$ with boundary $\partial M$ as:
\be\label{ec}
I_{E}= -\f{1}{16\pi G} \int_{M} (R-2\Lambda ) -
\f{1}{8\pi G} \int_{\partial M} K ,
\ee
which, upon variation, gives the vacuum Einstein equations with (++...+)
signature:
\be\label{ea1}
R_{AB} - \f{1}{2} g_{AB}R +\Lambda g_{AB} =0.
\ee

In Euclidean quantum gravity, the central object is the partition
function, Z, defined by the path integral over all metrics:
\be
Z=\int D[g] e^{-I_{E}[g]}.
\ee
In a semiclassical approximation, the dominant contribution to $Z$ will
come from fluctuations around a saddle point of the action. However, the
Euclidean action (\ref{ec})  diverges for non-compact geometries, since
the Einstein equations imply that $R$ is constant so the integral is just
a multiple of the infinite volume of the space. Furthermore, one should
also impose some  boundary conditions to fluctuations around
such spaces to define the path integral properly. To
have a well defined $Z$ then, one can pick up a background geometry
$\bar{g}_{AB}$, obeying (\ref{ea1}),  and use a modified action
\be\label{5}
\tilde{I}_{E}[g]= I_{E}[g] - I_{E}[\bar{g}]
\ee
in the definition of the partition function:
\be\label{666}
\tilde{Z}=\int D[g] e^{-\tilde{I}_{E}[g]}.
\ee
This new partition function $\tilde{Z}$ is related to $Z$ by  a
normalization with 
an infinite constant. If one includes in the path integral the geometries
which approach the background geometry at infinity such that 
$\tilde{I}_{E}[g]$ is finite, then $\tilde{Z}$ becomes well defined.
For a saddle point, using Einstein equations and up to a finite constant 
\footnote{Like in \cite{hp}, (\ref{5}) and (\ref{ma}) may not be equal
when the  integration regions of the
background and the saddle point are different. In this case
an additional but finite contribution should be added to (\ref{ma}).
However, this  point will not alter our discussion
of boundary conditions for which we will mainly use (\ref{ma}).},
$\tilde{I}_{E}$ may be rewritten as:
\be\label{ma}
\tilde{I}_{E}=-\f{1}{8\pi G}\int_{\partial M}  (K - \bar{K}),
\ee
where $\bar{K}$ is the trace of the extrinsic curvature of the boundary
embedded in the background space.\\

In our case the background space is the $d$-dimensional AdS soliton which
is given by \cite{hm}(in Euclidean signature):
\be\label{adssoliton}
ds^{2}=a^{2}r^{2}\left[ \left( 1-\f{r_{0}^{p+1}}{r^{p+1}}\right) d\tau^{2}
+ (dx^{i})^{2}\right] +
\f{1}{a^{2}r^{2}}\left( 1-\f{r_{0}^{p+1}}{r^{p+1}}\right) ^{-1}dr^{2},
\ee
where $r$ is a radial coordinate restricted to $r>r_{0}$, $\tau$ is an
angular coordinate, $x^{i}$ (with $i=1...p$ and $d=p+2$)  are coordinates
on $p$-dimensional Euclidean space
$R^{p}$ and $a$ is the inverse radius which is related to cosmological
constant by $a^{2}=-2\Lambda / (d-1)(d-2) $. To avoid a conical
singularity at $r=r_{0}$ one should identify $\tau$ with period $\beta =
4\pi / (a^{2} (p+1) r_{0}) $. We also define $l^{2}=a^{2} (d-1)$ so that
the Einstein equations (\ref{ea1}) may be written as:
\be\label{ea}
R_{AB} = -l^{2} g_{AB}.
\ee
The $p=3$ solution  is related to the Yang-Mills
gauge theory in 4-dimensions with one compactified direction. On the other
hand, the $p=5$ case is related to the exotic 6-dimensional world-volume 
theory of coincident M-theory fivebranes with one compact direction,
which reduces to Yang-Mills gauge theory in four-dimensions by a further
compactification.\\

In using this form of the metric in a semiclassical analysis, one faces
two important technical problems. The first one is that the boundary at
$r\to\infty$ of AdS soliton is $R^{p}\times S^{1}$.
Therefore, the modified action (\ref{ma}) may also diverge even for
suitably well defined boundary conditions due to the
integral over $R^{p}$. The second closely related problem is that the
interior of AdS soliton (regions of finite $r$) is also non-compact which 
implies the existence of  other boundaries in addition to boundary
at $r\to\infty$ (like the boundary at
$x^{i}x^{i}\to\infty$). One should also impose certain 
boundary conditions along these hypersurfaces.\\

There are different ways of overcoming these technical problems. One may
further redefine the partition function (\ref{666}) by replacing
$\tilde{I}_{E}$ by  $\tilde{I}_{E}/\cal{V}$, where $\cal{V}$  is the
volume of the space spanned by the coordinates $x^{i}$. Then, one should
impose the boundary conditions that $\partial_{i}g_{AB}=0$ when
$x^{i}x^{i}$ becomes greater then a constant. This ensures that
$\tilde{I}_{E}/\cal{V}$ does not diverge, which in turn gives a well
defined partition function.\\

Instead of dealing with action densities, we will proceed by modifying the
AdS soliton by identifying  the coordinates $x^{i}$ 
with period $L$. This  modification alters the asymptotic boundary from
$R^{p}\times S^{1}$ to
$T^{p}\times S^{1}$. One should not then worry about the two problems
mentioned above: the asymptotic integrals are now over $T^{p}\times S^{1}$
which will be convergent for suitable boundary conditions 
at $r\to \infty$, and the interior region becomes compact; the only
boundary is now located at $r\to \infty$.  The original AdS soliton may be
recovered by letting $L$ go to infinity.\\

Asymptotic boundary conditions  will  be fixed by demanding that 
(\ref{ma}) be finite for an  instanton.
Before discussing these boundary conditions, let us point
out another physical motivation for this choice. There is a close
connection between (\ref{ma}) and the total energy
defined in the Lorentzian context. 
Suppose we find an instanton  solution,
which asymptotically becomes the AdS soliton, and try to
determine to which Lorentzian  spacetime the AdS soliton decays (if it
does). To find this spacetime, one should make an inverse analytical
continuation of the instanton \cite{w2}.
Assuming that  $x^{1}= 0$ is a plane of symmetry, this may be achieved by
sending $x^{1} \to i x^{1}$.  
The coordinate $x^{1}$ now plays the role of time and  the energy of the
Lorentzian solution, calculated  with respect to the AdS soliton, turns
out to be finite, when(\ref{ma}) being evaluated on  the instanton is
finite. This may easily be seen from the definition  of total energy given
in \cite{hh} and shows that the physically interesting instantons have
$finite$ $actions$. Note that a similar relation between the value of the
Euclidean action and energy is also encountered in the asymptotically
flat context.\\

Let us now turn to the determination of the asymptotic behavior of 
a possible instanton solution. It
is assumed that outside a compact region, one can introduce angular 
coordinates $x^{\al} \equiv \tau , x^{i}$ (with periods $\beta$ and $L$
respectively) and $r$ in which the metric 
becomes asymptotically the AdS soliton:
\be
g_{AB}=\bar{g}_{AB} + h_{AB}, \hs{3}h_{AB}\to 0 \hs{3}as\hs{2} r\to \infty
,
\ee
such that (\ref{ma}) is finite. We will make no assumptions about the
topology of the compact region. In the asymptotic region and to the 
leading order in $h_{AB}$,  the inverse of the metric can be found as:
\be 
g^{AB}=\bar{g}^{AB} - \bar{g}^{AC}h_{CD}\bar{g}^{DB}.
\ee
The volume form on the boundary grows like $r^{p+1}$, therefore, to have a
well defined Euclidean action  we should demand that at large $r$,
$K-\bar{K}=O(1/ r^{p+1})$. Since the boundary is a $r=constant$
hypersurface, the unit normal vector, $n_{A}$, to this surface is given
by:
\be\label{unit}
n_{A}=\f{1}{\sqrt{g^{rr}}}\delta^{r}_{A}.
\ee
Using $K=\nabla_{A}n^{A}$ and further assuming  
that the $r$ derivatives of the metric fall off one
power faster, the finiteness of (\ref{ma}) implies for large $r$:
\footnote{ If we did not periodically identify the coordinates $x^{i}$,
then we had to impose the boundary conditions $\partial_{i}g_{AB}=0$ when
$x^{i}x^{i}$ is greater then a constant.} 
\be\label{metric1}
h_{\al\bt}=O(\f{1}{r^{p-1}}), \hs{5}h_{\al r}=O( \f{1}{r^{p}}),
\hs{5}h_{rr}=O( \f{1}{r^{p+3}}).
\ee
One can also calculate the asymptotic form of the Christoffel symbol as:
\be
\G^{r}{}_{\al\bt} = -a^{2}r^{3}\delta_{\al\bt} + O(\f{1}{r^{p-2}}),
\hs{5}
\G^{r}{}_{\al r} = O(\f{1}{r^{p-1}}),
\ee
\be
\G^{r}{}_{rr} = -\f{1}{r} + O(\f{1}{r^{p+2}}),
\hs{5}
\G^{\al}{}_{\bt\delta} =  O(\f{1}{r^{p-1}}), 
\ee
\be\label{16}
\G^{\al}{}_{\bt r} = \f{1}{r}\delta^{\al}_{\bt} + O(\f{1}{r^{p+2}}), 
\hs{5}
\G^{\al}{}_{rr} = O(\f{1}{r^{p+3}}).
\ee
As it was pointed out in \cite{HT}, the perturbation $h_{\alpha r}$ can be
made to fall off faster than $1/r^{p}$ by an appropriate choice of
coordinates at infinity.\footnote{I thank to M. Henneaux for pointing this
out to me.} Since this will not change the main conclusions of the paper,
we do not fix this gauge freedom at this point. In the same paper, the
physical boundary conditions were determined for a Lorentzian metric
approaching  asymptotically to $AdS_{4}$.\\

Compared to the asymptotically flat instanton like solutions \cite{w2},
the metric approaches the background geometry more rapidly. We will use
(\ref{metric1})-(\ref{16}) at several different points. 
In most cases these will  be  sufficient for us to obtain the results  we
are seeking. On the other hand, at one important instance when we try to
establish the existence of Killing vectors, we will be forced to consider
more restrictive boundary conditions.


\section{Hermitian differential operators and zero modes}


We are searching properties of possible Euclidean metric solutions of
(\ref{ea}) obeying the asymptotic boundary conditions determined in the
previous section. Let us denote one of such solutions by $M$. In this
section, we will introduce and study certain differential operators and
mainly try to see if they are invertible on suitable Hilbert
spaces.


\subsection*{\bf Operators $L_{1}$ and $L_{2}$}


We define the following differential operators on the tangent
bundle of $M$:
\be
L_{1}k_{A}\equiv \nabla^{2}k_{A} - l^{2} k_{A}, 
\ee 
\be
L_{2}k_{A}\equiv \nabla^{2} k_{A} + l^{2} k_{A}, 
\ee 
which will  act on the divergence free vector fields,
\be\label{div}
\nabla_{A}k^{A} = 0.
\ee
The reason for introducing these operators will become clear as we
proceed.
The vectors satisfying (\ref{div}) is a subspace of the total space of
vector fields and will be denoted by $V$. Let us first  show that $L_{1}$
and $L_{2}$ are well defined operators on this subspace:
\bea 
L_{1} : V \to V,\\ 
L_{2} : V \to V.
\eea
To verify this, we start from the curvature identity
\be
(\nabla_{A}\nabla_{B}- \nabla_{B}\nabla_{A})\nabla_{C}k_{D}=
R_{ABC}{}^{E}\nabla_{E}k_{D} + R_{ABD}{}^{E}\nabla_{C}k_{E},
\ee
and contract this with $g^{BC}$ and $g^{AD}$ to obtain
\be\label{eq1}
\nabla^{A}\nabla^{2}k_{A}-\nabla^{B}\n^{A}\n_{B}k_{A}=0.
\ee
Now, another curvature identity together with the Einstein equations
(\ref{ea}) give:
\be\label{eq2}
\n^{A}\n_{B}k_{A}=\n_{B}\n^{A}k_{A} - l^{2} k_{B}.
\ee
Using this in the second term of the (\ref{eq1}), one obtains  (for an
arbitrary vector field $k_{A}$),
\be
\n^{A}\n^{2}k_{A} = \n^{2}\n^{A}k_{A} - l^{2}\n^{B}k_{B}.
\ee
If $k_{A} \in V$ then $\n^{A}k_{A}=0$ and the above equation gives
 $\n^{A}\n^{2}k_{A}=0$. In this case, the definitions of $L_{1}$ and
$L_{2}$ imply:
\be
\nabla^{A} (L_{1}k_{A}) = 0,
\ee
\be
\nabla^{A} (L_{2}k_{A}) = 0,
\ee
showing that $L_{1}k_{A}\in V $ and $L_{2}k_{A}\in V$ when $k_{A}\in V$.
This verifies that
$L_{1}$ and $L_{2}$ are well defined operators on $V$.\\

We introduce the following inner product on the vector space $V$:
\be\label{inp}
<k|l>=\int_{M} g^{AB}k_{A}l_{B}.
\ee
In general, there are vectors in $V$ which do not have well defined inner
products. Thus, some boundary conditions should be imposed to 
eliminate these vector fields. 
The fact that, when $r$ is large,  the volume form on $M$ grows like
$r^{p}$  and the asymptotic behavior  of the metric fixed in the
previous section suggest the following boundary 
conditions on the components of $k_{A}$ (in $x^{\al},r$ coordinates):
\be\label{bc}
k_{r}=O(\f{1}{r^{p/2 + 2}}),\hs{5} 
k_{\al}=O(\f{1}{r^{p/2}}).
\ee
We $redefine$ $V$ to be the space of divergence free vector fields obeying
these boundary conditions. With this redefinition, the
inner product (\ref{inp}) becomes  well defined on $V$. Note that $V$ is
the kernel of the elliptic operator $*d*$ acting on the
one-forms. Due to this fact, we assume that $V$ satisfies the
axioms of Hilbert spaces rigorously. We also assume that
the $r$-derivatives of the components fall off one power faster.\\

Let us show that $L_{1}$ and $L_{2}$ are Hermitian operators
on $V$. This requires vanishing of the following surface integral,
\be\label{st}
\int_{\partial M} n^{A}k^{B}(\nabla_{A}l_{B})=0,
\ee
for $all$ vectors $k_{A},l_{A}\in V$, where $n^{A}$ is the unit normal to
$\partial M$. The volume form on $\partial M$ grows like $r^{p+1}$ at
large $r$. Using the expression (\ref{unit}) for the unit normal vector,
the asymptotic behavior of the metric and the 
boundary conditions (\ref{bc})
obeyed by the vector fields $k_{A}$ and $l_{A}$, one
can check that the integrand vanishes as $r\to\infty$. 
Then, (\ref{st}) implies, upon integration by parts,
\be 
\int_{M} k^{A}\nabla^{2}l_{A} =  \int_{M} l^{A}\nabla^{2}k_{A},
\ee 
which in turn proves that $L_{1}$ and $L_{2}$ are Hermitian operators on
$V$:
\bea
<k|L_{1}l>= <L_{1}k|l>,\\
<k|L_{2}l>= <L_{2}k|l>.
\eea
The boundary conditions which are imposed to obtain a well defined inner
product turn out to be sufficient to make $L_{1}$ and $L_{2}$
Hermitian.\\
 
We will now determine the asymptotic behavior of a vector field $k_{A}\in
V$ which  satisfies $L_{1}k_{A}=0$ or $L_{2}k_{A}=0$.
We  start from the fact that, when $r$ is large  the components  of
$k_{A}$ become:
\be
k_{r}=O(\f{1}{r^{m}}),\hs{5}k_{\al}=O(\f{1}{r^{n}}),\hs{5} m\geq\f{p}{2}
+2, \hs{5}n\geq\f{p}{2}.
\ee
To find the possible values of $m$ and $n$, one can expand the components
in powers of $1/r$ (which can be done when $r$ is
sufficiently large):  
\bea\label{ex} 
k_{r}=\f{f(x)}{r^{m}} + \f{g(x)}{r^{m+1}} + ...,\\
k_{\al} = \f{f_{\al}(x)}{r^{n}} + \f{g_{\al}(x)}{r^{n+1}}+.... 
\eea 
After plugging the above expansions into 
the differential equations, one can group
the terms according to their powers of $1/r$ and solve
equations order by order.
The lowest order equations, which are  sometimes called the indicial
equations,  determine the
constants $m$ and $n$. At large $r$, the components of $\nabla^{2}k_{A}$
can be calculated as (we set the inverse radius $a=1$ from now on):
\bea\label{l1} 
\nabla^{2}k_{r}=
r^{2}\partial_{r}^{2} k_{r} + \f{1}{r^{2}}
\delta^{\al\bt}\partial_{\al}\partial_{\bt} k_{r} + k_{r} + (p+4) r
\partial_{r}k_{r} - \f{2}{r^{3}}
\delta^{\al\bt}\partial_{\al}k_{\bt}\\ \nonumber 
+ O(\f{1}{r^{p+m+1}})+O(\f{1}{r^{p+n+2}}),
\eea 
\bea\label{l2} 
\nabla^{2}k_{\al}= r^{2}\partial_{r}^{2} k_{\al} +
\f{1}{r^{2}} \delta^{\gamma\bt}\partial_{\gamma}\partial_{\bt} k_{\al} -
(p+1)  k_{\al}+ p r \partial_{r}k_{\al} + 2 r \partial_{\al}k_{r}
\\ \nonumber + O(\f{1}{r^{p+n+1}})+ O(\f{1}{r^{p+m-2}}),
\eea 
where the suppressed  terms, which arise due to the deviation of the
metric  from the AdS soliton background, will not affect the indicial
equation. The terms
$\f{1}{r^{2}}\delta^{\al\bt}\partial_{\al}\partial_{\bt} k_{r}$
and
$\f{1}{r^{2}}\delta^{\gamma\bt}\partial_{\gamma}\partial_{\bt}k_{\al}$, 
appeared in (\ref{l1}) and (\ref{l2}),  will only contribute to 
the higher order equations,
even though they survive when the deviation
from the background is zero. Thus, they can also be ignored in finding the
possible values of $m$ and $n$. 
The fact  that the vector fields are divergence free,
$\nabla_{A}k^{A}=0$, gives the following relation among the components: 
\be \label{dv}
r^{2}\partial_{r}k_{r}+(p+2)rk_{r}+\f{1}{r^{2}}
\delta^{\al\bt}\partial_{\al} k_{\bt} + 
O(\f{1}{r^{p+m}})+ O(\f{1}{r^{p+n+1}})=0, 
\ee 
where the suppressed  terms have higher powers of $1/r$ compared to the
ones that are written. Now, using (\ref{dv}) for the last term of
(\ref{l1}), one can show that the lowest order $1/r$ equations, obtained
from the  $r$-components of the differential
equations $L_{1}k_{A}=0$ and $L_{2}k_{A}=0$, 
give the following quadratic equations for $m$: 
\be\label{m1}
L_{1}k_{A}=0\hs{5}\Rightarrow\hs{5} m^{2} - (p+3) m + (p+4) =0, 
\ee
\be\label{m2}
L_{2}k_{A}=0 \hs{5}\Rightarrow \hs{5}m^{2} - (p+3) m + (3p + 6) =0. 
\ee
Fortunately, all the terms which has the constant $n$ turn out to be
higher order and we finally end up with equations for $m$.\\

For the vector fields satisfying  $L_{1}k_{A}=0$, 
the constant  $m$ will have the following values.
\be
m_{1}=1,\hs{5} m_{2}=p+4.
\ee
The first root is not acceptable since it is not consistent with the
boundary conditions imposed on the vector fields. Thus, we set $m=p+4$.
Using 
(\ref{l2})
in the $\al$-component of the equation $L_{1}k_{A}=0$, one can
obtain a quadratic equation for $n$, 
\be
n^{2} - (p-1) n - (2p+2) =0,
\ee
which has the single positive root $n=p+1$. Therefore,  the solutions
$L_{1}k_{A}=0$, $k_{A}\in V$, fall off like:
\be\label{f1} 
k_{r}=O(\f{1}{r^{p+4}}),\hs{5} k_{\al}=O(\f{1}{r^{p+1}}).
\ee

For the vector fields obeying $L_{2}k_{A}=0$, the roots of $m$ can be read
off from (\ref{m2}) as:
\be
m_{1}=3, \hs{5} m_{2}=p+2.
\ee
The first root is not acceptable for the cases $p>1$, since it is not 
consistent with the boundary conditions,  and the roots become
equal for $p=1$. Therefore, we set $m=p+2$.
Then, (\ref{l2}) can be used in the  $\al$-component of $L_{2}k_{A}=0$, to
obtain the  following quadratic equation for $n$:
\be
n^{2}-(p-1) n=0.
\ee
This  has the single positive root $n=p-1$ and therefore
the solutions of $L_{2}k_{A}=0$, $k_{A}\in V$, become for
large $r$ 
\be\label{f2}
k_{r}=O(\f{1}{r^{p+2}}), \hs{5} k_{\al}=O(\f{1}{r^{p-1}}).
\ee
Till now we have fixed the boundary conditions which ensure that
the operators $L_{1}$ and $L_{2}$ are Hermitian
and determined the asymptotic behavior  of their zero modes. 
We will now try to obtain further information about the zero modes, since
our final aim is to use these operators in solving certain Dirichlet
problems which will possibly imply  existence of the Killing vectors. It
would be nice if our operators $L_{1}$ and $L_{2}$ turned out to be
invertible i.e. they did not have any zero modes. This will be the case
for $L_{1}$, and for $L_{2}$ we will be able to determine all the zero
modes.\\ 

We proceed  by showing that the Hermitian differential operator
$L_{1}$ does not have any zero modes. When $k_{A}\in V$,
the fact that its divergence free and  (\ref{eq2})  give 
\be\label{im}
\nabla^{A}\nabla_{B}k_{A}=-l^{2}k_{B},
\ee
which can be used to rewrite the equation  $L_{1}k_{A}=0$ 
as:
\be
\nabla^{A}(\nabla_{A}k_{B}+\nabla_{B}k_{A})=0.
\ee
We contract the above equation with $k^{B}$ and integrate over all $M$:
\be
\int_{M}k^{B}\nabla^{A}(\nabla_{A}k_{B}+\nabla_{B}k_{A}) = 0.
\ee
Integrating by parts, one can easily obtain, for any vector field
satisfying $L_{1}k_{A}=0$ and $\nabla_{A}k^{A}=0$,
\be\label{51}
-\f{1}{2}\int_{M} (\nabla^{A}k^{B}+\nabla^{B}k^{A})(\nabla_{A}k_{B} 
+ \nabla_{B}k_{A}) + \int_{\partial M} n^{A}k^{B}(\nabla_{A}k_{B} +
\nabla_{B} k_{A}) =0.
\ee
Now, the vector
fields in $V$ fall off like (\ref{f1}) when they obey $L_{1}k_{A}=0$.
This ensures that the surface integrals  vanish which in turn implies,
\be\label{rel}
\nabla_{A}k_{B} + \nabla_{B}k_{A} =0,
\ee
since the remaining integrand in (\ref{51}) is nowhere negative.
Therefore, $k_{A}$ is a Killing vector. By the asymptotics 
(\ref{f1}), both $k_{A}$ and $\nabla_{A}k_{B}$ vanishes on $\partial M$.
However, it is very well known that 
if a Killing field and its derivative vanish at any point, then
the Killing field vanishes everywhere \cite{wald}. Thus $k_{A}=0$ and
$L_{1}$ has no zero modes with the given boundary conditions.\\

Let us now try to learn about the zero modes of the operator $L_{2}$. We
will simply repeat the above steps, to obtain a relation like
(\ref{rel}). Using (\ref{im}), the equation $L_{2}k_{A}=0$ can be
rewritten as,
\be
\nabla^{A}(\nabla_{A}k_{B}-\nabla_{B}k_{A})=0.
\ee
We first  contract the above equation with $k^{B}$ and integrate over
all $M$. Then, integrating by parts, we obtain for any vector field
satisfying $L_{2}k_{A}=0$ and $\nabla_{A}k^{A}=0$: 
\be\label{54} -\f{1}{2} 
\int_{M}(\nabla^{A}k^{B}-\nabla^{B}k^{A})(\nabla_{A}k_{B}-\nabla_{B}
k_{A}) + \int_{\partial M} n^{A}k^{B}(\nabla_{A}k_{B}-\nabla_{B}k_{A})
=0.   
\ee
The surface integrals vanish due to the asymptotic behavior
(\ref{f2}). The remaining integrand is positive definite which 
implies that 
\be 
\nabla_{A}k_{B}-\nabla_{B}k_{A}=0.
\ee
Therefore the zero modes of the Hermitian operator $L_{2}$ are closed
one-forms. Since $\nabla_{A}k^{A}=0$ they are also co-closed and thus
harmonic one-forms.\\

One can obtain further information about these zero modes
by recalling that  a closed one-form 
can be written as a linear combination of an exact one-form and first
cohomology classes of the manifold $M$. The cohomology classes are related
only to the topology of the manifold. In our case, we have assumed that,
outside  a compact region, the topology of $M$  become globally
the AdS soliton i.e.  $T^{p}\times S^{1}$ times a real line (the
coordinate $r$). Therefore, near the asymptotic region,  the first
cohomology classes are spanned at most \footnote{In the interior region
the topology may change such that some of the one-forms $dx^{\al}$ become
also exact. This is the case, for instance, for the one-form $d\tau$ in
the solution (\ref{adssoliton}).}
by the one-forms $dx^{\al}$. However, these
forms cannot be the zero modes of $L_{2}$ since they do not obey the
boundary conditions imposed on the vector fields belonging to $V$.
We will show in a moment that the exact forms should also be excluded. 
Therefore, the zero modes of $L_{2}$ are the harmonic representatives
of first cohomology classes of the interior compact region. \\

To see that the exact forms cannot be
the zero modes of $L_{2}$ we write $k_{A}=\nabla_{A}f$, for some function
$f$. By the asymptotic behavior of the zero modes of $L_{2}$ determined
in (\ref{f2}), $f$ should behave for large $r$ like,
\be\label{func}
f=O(\f{1}{r^{p+1}}).
\ee 
Since $k_{A}$ is also  divergence free, $f$ should
satisfy the Laplace equation: $\nabla^{2}f=0.$
However, any solution of the Laplace equation which obeys 
(\ref{func}) should be zero. This can easily be seen by nothing 
the following relation which is valid for any function obeying Laplace
equation:
\be\label{lap}
0=\int_{M} f\nabla^{2}f=-\int_{M}(\nabla_{A}f)(\nabla^{A}f) +
\int_{\partial M} n^{A} f (\nabla_{A}f) .
\ee
The last surface integral vanishes due to the behavior  (\ref{func}) of
$f$ at large $r$. Since the remaining integrand in the right 
side is positive definite, one obtains $\nabla_{A}f=0$.
Therefore, $f$ is a constant and the constant vanishes since $f$ vanishes
at $r\to \infty$. This shows that  exact forms should be excluded from
the list of the zero modes of $L_{2}$.


\subsection*{\bf The Laplacian}


Having studied properties of the differential operators acting on
divergence free vector fields, we finally turn to the space of functions.
Let $W$ be the space of square integrable functions endowed with  the
following inner product,
\be
<f|g>=\int_{M} f g.
\ee
The inner product is well defined, since the functions in $W$ obey at
large $r$ (due to square integrability):
\be \label{bcforlap}
 f\in W \hs{5}\Rightarrow \hs{5} f=O(\f{1}{r^{p/2 + 1}}).
\ee
The differential operator that we will consider on $W$ is the Laplacian. 
One can easily check that, due to the asymptotic behavior of
functions determined above,  $\nabla^{2}$ is an Hermitian operator 
\be
<f|\nabla^{2}g>=<\nabla^{2}f|g>.
\ee
As before, we try to determine the zero modes of $\nabla^{2}$ and see if
it is invertible on $W$. The asymptotic behavior of the zero modes can
be determined by calculating 
\be\label{wlap}
\nabla^{2}f = r^{2}\partial_{r}^{2} f +
\f{1}{r^{2}}\delta^{\al\beta}\partial_{\al}\partial_{\beta} f - (p+2) r
\partial_{r} f + O(\f{1}{r^{p+m+1}}),
\ee
where $f=O(1/r^{m})$. 
The indicial equation which follows from the above formula reads
\be\label{62}
m(m+1) - (p+2) m = 0.
\ee
The only solution which is consistent with the boundary
condition (\ref{bcforlap}) is $m=p+1$. Therefore,  zero modes of
$\nabla^{2}$ in
$W$ fall off like $1/r^{p+1}$. However, repeating the above arguments 
which has been used to show that the exact forms cannot be the zero modes
of $L_{2}$, one can easily see that the zero modes should  actually be
zero. Therefore the Laplacian $\nabla^{2}$ is an invertible
operator on $W$. 


\section{Existence of harmonic functions  and commuting  Killing
vectors}


In the previous section we have considered two vector spaces $V$ and $W$,
endowed with well defined inner products, and studied properties of the 
differential operators $L_{1}$, $L_{2}$ and $\nabla^{2}$. We have shown
that all operators are Hermitian and determined the zero modes.
The operators $L_{1}$ and $\nabla^{2}$ are shown to have no zero modes
and thus are invertible. 
In this section we will try to use these operators to show the
existence of harmonic functions and Killing vectors.


\subsection*{\bf Existence of harmonic functions}


Although $\nabla^{2}$ has no zero modes in $W$, there may be solutions 
of $\nabla^{2} f=0$, where $f \not\in W$. Consider the  following
Dirichlet problem 
\be\label{dirichlet2}
\nabla^{2}f=0, \hs{5} f\to b_{i}x^{i} \hs{5} as \hs{5} r\to \infty ,
\ee
where $b_{i}$ are arbitrary constants. Note that the index $i$ runs from 1
to $d-2$. Since the coordinates $x^{i}$ are periodically identified
with the period $L$, the function $f$ is not well defined for finite $L$.
However, the constant $L$ is a free parameter  and one
can let $L\to \infty$. Indeed, as discussed in section 2, this limit 
corresponds to the AdS soliton. On the other hand one cannot
include the coordinate $\tau$ in (\ref{dirichlet2}) since it's period
$\beta$ is fixed ($\beta$ corresponds to either compactification scale or
the inverse temperature of the boundary conformal field theory).
Note also that $f \not\in W$.\\ 

To solve the above Dirichlet problem we first define a trial
function $t$ such that near the asymptotic boundary  (where the
coordinates $x^{\alpha}$ and $r$ are well defined) it becomes
$t=b_{i}x^{i}$. We smoothly extend $t$ to be zero through the 
interior regions (where the coordinates $x^{\alpha}$ and $r$ may not be
well defined) and thus obtain a well defined function on $M$.
We find the  solution of (\ref{dirichlet2}) by writing
$f=l + t$, and solving for $l$ obeying
\be\label{77}
\nabla^{2}l = -\nabla^{2} t,\hs{5} l\to 0 \hs{5} as \hs{5} r \to\ \infty .
\ee
Using (\ref{wlap}), one can check  that $\nabla^{2}t=O(1/r^{p+2})
\in W$ 
\footnote{One may naively claim from (\ref{wlap}) (by setting $m=0$)
that $\nabla^{2}t = O(1/r^{p+1})$. However, a carefull
analysis for $t$ shows that $O(1/r^{p+1})$ terms cancel each
other in (\ref{wlap}) which implies $\nabla^{2}t=O(1/r^{p+2})$.}
and thus there is a unique solution for $l \in W$, since $\nabla^{2}$ is
invertible. By (\ref{bcforlap}), $l=O(1/r^{p/2+1})$. To obtain more
accurate information about the asymptotic behavior of $l$, we expand it
at large $r$ as 
\be
l=\frac{f(x)}{r^{m}}+\frac{g(x)}{r^{m+1}}+.... \hs{10} m\geq
\frac{p}{2}+1.
\ee
Now, (\ref{wlap}) implies that  
the lowest order terms  of $\nabla^{2}l$ fall off like 
$1/r^{m}$. Since $l$ obeys (\ref{77}), these terms should either cancel
each other or be equal to the  same order terms coming from  the right
hand side. Thus, either $m$ should obey the indicial equation (\ref{62})
and $m=p+1$ or $m=p+2$. 
This implies, at least, $l=O(1/r^{p+1})$ which, furthermore, gives the
following solution to the Dirichlet problem (\ref{dirichlet2}):
\be\label{clv}
f=b_{i}x^{i} + O(\f{1}{r^{p+1}}).
\ee
One can also try to solve for $l$ by constructing a Green's function,
$G(r,r')$,  for $\nabla^{2}$. The asymptotic behavior of $G(r,r')$  can be
deduced from the asymptotics of the possible zero modes since it
obeys the Laplace equation except at some singular points. 
Inverting (\ref{77}) as $l(r)=-\int G(r,r')\nabla^{2}t(r')$ and letting
$r\to \infty$, one finds that $l=O(1/r^{p+1})$ and thus $f$ has the same
asymptotic behavior (\ref{clv}).  
Since $f$ obeys $\nabla^{2}f=0$, it is an harmonic function
$d*df=0$. Note also that the one-form $df$ is also harmonic
$(d\delta + \delta d)df=0$, where $\delta$ is the adjoint of $d$.
On the other hand, since in the construction we have introduced 
arbitrary constants $b_{i}$, there are actually $d-2$ independent
harmonic functions. We will denote the harmonic function
obtained by the choice $b=(0,0,..1,0..0)$ by $f_{(i)}$, where in $b$ only
the i'th entry is one. All these are valid for $all$  instanton
solutions.


\subsection*{\bf Existence of Killing vectors}


The  Hermitian operator $L_{1}$ is invertible on $V$. However,
there may be solutions of $L_{1}k_{A}=0$, where $k_{A}$ is a 
divergence free vector field which does not belong to $V$. Let us try to
see if the following Dirichlet problem has a solution,
\bea\nonumber
L_{1}k_{A}=0, \hs{5}\nabla_{A}k^{A}=0
\eea
\be\label{dic}
k^{\al}\to a^{\al}, \hs{5}
k^{r}\to 0, \hs{5} as \hs{5} r\to \infty ,
\ee
where  $a^{\al}$ are arbitrary constants. Note that, the 
$contravariant$ components of the vector field are fixed on the boundary.
On the other hand, the covariant components become, for example,
$k_{\al}=O(r^{2})$  which  shows that $k_{A}\not\in V.$ \\
   
To solve this problem, we first define a trial vector field
$t_{A}$ with the following properties. Near the asymptotic boundary,
the components $t_{\al}(x,r)$
are smooth functions of coordinates $x^{\al},r$ such that they vanish
when $r<R_{-}$ and become exactly $a^{\al}r^{2}$
when $r>R_{+}$, where $R_{-}$ and $R_{+}$ are sufficiently large numbers
with $R_{-}<R_{+}$.  The component function,
$t_{r}(x,r)$, is defined to be zero when $r<R_{-}$ and
will be chosen to give $\nabla_{A}t^{A}=0$ when $r>R_{-}$. We extend
$t_{A}$ to be zero trough the interior
regions, where the coordinates $x^{\al}$ and $r$
may not be well defined. This gives us a well defined divergence free
vector field over all $M$.\\

We introduce a new vector field $l_{A}$ by writing the desired solution of 
(\ref{dic}) as:
\be
k_{A}=l_{A}+t_{A}.
\ee
By the construction of the trial vector field $t_{A}$, one can 
solve the Dirichlet problem (\ref{dic}) by working the solutions
$l_{A}$ of,
\be\label{22}
L_{1}l_{A}=-L_{1}t_{A}, \hs{5} l_{A}\to 0,\hs{5} as\hs{5} r\to\infty .
\ee
We will show that $L_{1}t_{A}\in V$ when $p>3$.
Since $L_{1}$ is an
invertible operator on $V$, there is a unique solution of (\ref{22})
with $l_{A}\in V$.  This will give a unique solution of the Dirichlet
problem (\ref{dic}) for $p>3$.\\

The $\al$-component $t_{\al}$
of the trial vector field is fixed by construction. Let us
determine the behavior of $t_{r}$ at large $r$. As mentioned above,
$t_{r}$ is chosen to obtain a divergence free vector field, thus
satisfies a first order differential equation.  Using 
(\ref{dv}) for the components of $t_{A}$, we obtain at large $r$
\be
t_{r}=O(\f{1}{r^{p}}). 
\ee
Using (\ref{l1}) and (\ref{l2}), one can see that, in
$L_{1}t_{A}$, the terms which come from the background and 
contain the constants $a^{\al}$ cancel
each other. The remaining terms can be determined to fall off like
\be\label{tf}
(L_{1}t)_{r} = O(\f{1}{r^{p}}),\hs{5} (L_{1}t)_{\al}=O(\f{1}{r^{p-1}}).
\ee
This shows that $L_{1}t_{A}\in V$ when $p>3$,  since it obeys the
boundary conditions (\ref{bc}) only on that range. As mentioned above,
$L_{1}$ can be inverted to obtain a unique
solution $l_{A}\in V$. Therefore, the components of $l_{A}$ obey at least
(\ref{bc}). However, we will need more accurate information about the
asymptotic behavior of $l_{A}$. To obtain this information, 
we start from the fact that at large $r$:
\be
l_{r}=O(\f{1}{r^{m}}),\hs{5}l_{\al}=O(\f{1}{r^{n}}),\hs{5} m\geq\f{p}{2}
+2, \hs{5}n\geq\f{p}{2},
\ee
and expand the components like in (\ref{ex}). The lowest order terms of
the $r$-component of $L_{1}l_{A}$ can be seen to  fall off like
$1/r^{m}$. At this stage there are two possibilities; these terms
may cancel each other or they may be equal to the same order terms 
coming from the right hand  side of (\ref{22}) i.e. $-(L_{1}t)_{r}.$ 
For these terms to cancel each other, $m$ should satisfy exactly the same
indicial equation (\ref{m1}) obtained for the zero modes. 
If this is the case then $m=p+4$, but now in $(L_{1}l)_{r}$ there is no
term which can cancel the lowest order terms coming from  $(L_{1}t)_{r}$. 
Thus the second possibility should be true. The asymptotic behavior of
$(L_{1}t)_{r}$ implies that $m=p$ and $l_{r}=O(1/r^{p})$. The same
arguments can be repeated for the $\al$-components to obtain:
\be
l_{r}=O(\f{1}{r^{p}}),\hs{5}l_{\al}=O(\f{1}{r^{p-1}}).
\ee
This shows that there is a unique solution of $L_{1}k_{A}=0$,
$\nabla_{A}k^{A}=0$, given by
$k_{A}=l_{A}+t_{A}$, which asymptotically becomes (contravariant
components):
\be\label{kilv}
k^{r}=O(\f{1}{r^{p-2}}), \hs{5} k^{\al}=a^{\al} + O(\f{1}{r^{p+1}}).
\ee
Note that, to be able to solve the Dirichlet problem, we are forced to
impose $p>3$ i.e. the dimension of $M$ should be greater than 5.\\

We now try to establish that (\ref{kilv}) is a Killing
vector. To show this, one can use the equation (\ref{51}), which is valid
for any vector field obeying $L_{1}k_{A}=0$ and $\nabla_{A}k^{A}=0$. If
the surface integral in (\ref{51}) vanishes, then one has 
$\nabla_{A}k_{B}+\nabla_{B}k_{A}=0$, since the remaining terms  
in (\ref{51}) are positive definite.
There is $only$ $one$ contribution to the surface integral, 
coming from  $n^{r}k^{\al}(\nabla_{\al}k_{r}+\nabla_{r}k_{\al})$, 
which does not vanish. Since the volume form
of $\partial M$ grows like $r^{p+1}$, $n^{r}=O(r)$ and $k^{\al}=O(1)$, 
the surface integral integral of this term vanishes when
\be\label{bad}
\nabla_{\al}k_{r}  + \nabla_{r}k_{\al} = O(\f{1}{r^{p+3}}).
\ee
However, one can easily check that this condition is $not$ satisfied
by the solution (\ref{kilv}) of the Dirichlet problem. 
The solution (\ref{kilv}) obeys, instead, 
$\nabla_{\al}k_{r}+\nabla_{r}k_{\al} = O(1/r^{p})$, which shows that the
surface integral may not vanish in general. Note that
$a^{\al}\f{\partial}{\partial x^{\al}}$ is already an asymptotic Killing
vector field of any possible instanton like solution. To see on which
spaces the surface integral vanishes, let us  concentrate on a class of
instanton solutions obeying more restrictive boundary conditions such as
\be\label{newasy}
h_{\al\beta}=O(\f{1}{r^{p+2}}), \hs{5} h_{\al r}=O(\f{1}{r^{p+3}}),
\hs{5} 
h_{rr}=O(\f{1}{r^{p+6}}).
\ee
Compared to the asymptotic behavior (\ref{metric1}), the metrics of  this
family fall off three power of $r$ faster. In order to take into account
of this difference in the previous formulas, one should simply shift
$p$ by 3. This implies that the Dirichlet problem  (\ref{dic}) can now be
solved in $any$ dimensions and the solution (\ref{kilv}) becomes
asymptotically  
\be\label{kilv2}
k^{r}=O(\f{1}{r^{p+1}}), \hs{5} k^{\al}=a^{\al} + O(\f{1}{r^{p+4}}).
\ee
By plugging (\ref{kilv2}) into the identity (\ref{51}), one can see that
the surface integrals do vanish. Therefore, (\ref{kilv2}) is indeed a
Killing vector. On the othar hand, since  the constants
$a^{\al}$ are  completely arbitrary, there are  actually $p+1$ independent
Killing vectors. We will denote the Killing vector field which is obtained
by the choice $a^{\al}=(0,0..1,..0)$ by $k^{(s)A}$ where $p+1\geq s\geq 1$
and in $a^{\al}$ only the $s$'th entry is one. We remind the reader that
the proof only applies to the family (\ref{newasy}).\\  

By defining 
\be 
z_{A} \equiv [k^{(s)},k^{(s')}]_{A}= 
k^{(s)B}\nabla_{B}k^{(s')}_{A} -  k^{(s')B}\nabla_{B}k^{(s)}_{A},
\ee
one can easily show that  the Killing vectors of this family also
commute. The asymptotic behavior (\ref{kilv2}) implies that
\be
z_{r}=O(\f{1}{r^{p+3}}),\hs{5} z_{\al}=O(\f{1}{r^{p+2}}).
\ee
Since the commutator of two Killing vectors is  again a Killing vector,
$z_{A}$ is a Killing vector. Being a Killing vector and having the above
asymptotic behavior, $z_{A}\in V$ and $L_{1}z_{A}=0$. However,
since $L_{1}$ is invertible, $z_{A}=0$. Thus the Killing
vectors commute with each other.


\subsection*{\bf Operator $L_{2}$}


We have studied Dirichlet problems related
to the operators $L_{1}$ and $\nabla^{2}$. As we have discussed, the
existence of the solutions imply the existence of certain geometric
properties. There is another operator, $L_{2}$, which we have studied in
section 2 and not considered yet in the context of Dirichlet problems.
Since the Hermitian differential operator $L_{2}$ is not invertible
on $V$, one cannot naively solve a Dirichlet problem 
using $L_{2}$ (at least following the steps discussed so far). 
It seems that one should also make some topological
assumptions to obtain an invertible operator. Here, we simply note that
the one-forms $\nabla_{A}f_{(i)}$ are divergence free and they are also
eigenvectors of $L_{2}$ with zero eigenvalue, 
$L_{2}\nabla_{A}f_{(i)}=0$. Note also that $\nabla_{A}f_{(i)}\not \in
V$.


\subsection*{Relation between $k_{A}^{(i)}$ and $df_{(i)}$}


We conclude this section by showing that all but one of the Killing
vectors are duals of harmonic one-forms $df_{(i)}$:
\be\label{dual}
k^{(i)A}\nabla_{A}f_{(i')}=\delta^{i}_{i'}.
\ee
Note that at $r=\infty$ the Killing vector $k^{(i)A}$ 
and the function $f_{(i)}$ equal to $\f{\partial}{\partial x^{i}}$ and  
$x^{i}$, respectively. To prove (\ref{dual}), using
$\nabla_{(A}k_{B)}^{(i)}=0$, $L_{1}k_{(i)A}=0$ and 
$L_{2}\nabla_{A}f_{(i)}=0$,  we calculate
\bea
\nabla^{2}\lf k^{(i)A}\nabla_{A}f_{(i')}\rg  = \lf\nabla^{2}k^{(i)A}\rg
\nabla_{A}f_{(i')} + k^{(i)A}\lf \nabla^{2}\nabla_{A}f_{(i')}\rg + 
\nonumber \\
2 \lf\nabla^{B}k^{(i)A}\rg \lf \nabla_{B}\nabla_{A}f_{(i')}\rg =0,
\eea
which shows that the functions $k^{(i)A}\nabla_{A}f_{(i')}$ obey the
Laplace
equation. One can now use (\ref{lap})  which is valid for any function
obeying Laplace equation. We focus on the surface integrals. If one writes
$g=k^{(i)A}\nabla_{A}f_{(i')}$, then using the asymptotic behavior 
(\ref{kilv2}) and (\ref{clv}) of $k^{(i)A}$ and $f_{(i)}$, respectively,
one can obtain
\be
g = \delta^{i}_{i'} + O(\f{1}{r^{p+1}}).
\ee
Then, the surface integrals in (\ref{lap}), obtained after plugging $g$
into, can be written explicitly as:
\be
\delta_{i'}^{i}\int_{\partial M} n^{A}\nabla_{A}g + \int_{\partial M}
O(\f{1}{r^{p+1}}) n^{A}\nabla_{A}g .
\ee
The first term in the above integrals vanishes since it can be written as
the integral of $\nabla^{2}g$ over all $M$ which is zero, and the second
integral vanishes since the volume form on $\partial M$ grows like
$r^{p+1}$, $n^{A}=O(r)$, $\nabla_{A}g = O(1/r^{p+1})$ which imply that the
integrand vanishes as $r\to \infty$. Therefore, the surface
integrals vanish in (\ref{lap}). The remaining integrand in (\ref{lap}) is
positive definite which implies that $\nabla_{A}g=0$. Therefore, $g$ is
constant and the constant is $\delta^{i}_{i'}$ since $g$ reaches this
value at $r\to\infty$. This proves (\ref{dual}), which in turn
shows that the Killing vectors $k_{A}^{(i)}$ and the one-forms $df_{(i)}$ 
cannot  $vanish$ at any point. Thus, all but one of the Killing vectors
do not have any fixed points and act on
$M$ freely.  This fact is important since it enables one to choose global
coordinates adapted to Killing vectors, which are the functions
$f_{(i)}$ in our case.

\section{Conclusions}

In this paper we have tried to obtain certain geometric properties of 
AdS instantons. After determining the asymptotic behavior of the metric, 
we have studied certain Hermitian differential operators, defined on
suitable Hilbert spaces. Using these
operators, we have shown that on a $d$-dimensional instanton there are
$d-2$ harmonic functions. For a class of instatons, we have also proved
the existence of $d-1$ commuting Killing vectors. Furhermore, all but
one of the Killing vectors are shown to have no fixed points.\\

These results may be usuful in proving the semiclassical
stability of AdS soliton. We believe that the existence proof of
the Killing vectors on the  family (\ref{newasy}) can be generalized to
all instanton solutions by choosing more
appropriate trial vector fields in solving the Dirichlet problem
corresponding to $L_{1}$. If this could be done,  one can introduce
globally well defined coordinates adapted to the Killing vectors
which act on the manifold freely, and  try to solve the Einstein
equations in these coordinates as an attempt to obtain all possible
instanton solutions.\\

The existence of an instanton solution does not necessarily imply 
that the AdS soliton is semiclassically  unstable. This solution may well
describe  the decay of another unstable state to AdS soliton. To see that
a ground state is semiclassically unstable, one should study the
quadratic fluctuations around an instanton solution. The negative action
modes which may arise in this quadratic approximation is a sign of
semiclassical instability.\\ 

In the the proofs of semiclassical stability of Minkowski space and the 
positive energy theorem for asymptotically flat spaces 
given in \cite{w1}, the Hermitian  operators 
and their Green's functions played the crucial role. The
operators introduced in this paper  may also be useful in studying the
positive energy conjecture of Horowitz and Myers. In the present paper,
the metric is positive definite and this plays the key  role at each
step in deriving the results. In the Lorentzian context, one can choose an
Euclidean
initial value hypersurface and try to adapt the arguments presented here
to obtain information about this Euclidean hypersurface. 

\subsection*{Acknowledgements}
I would like to thank S. Erdin and S. Gunturk for reading the manuscript.

\end{document}